\documentclass[11pt,a4paper]{article}
\usepackage[T1]{fontenc}
\usepackage{lmodern,amsmath,amssymb,amsthm,mathtools,mathrsfs}
\numberwithin{equation}{section}
\usepackage[margin=26mm]{geometry}
\usepackage{microtype}
\usepackage{color}
\usepackage{amsmath}
\usepackage{amssymb}
\usepackage[hidelinks]{hyperref}
\hypersetup{pdftitle={A direct Fock-space proof of the domain identity},
 pdfsubject={Finite-mode quadratic contraction generators and their domains}}
\usepackage{enumitem}
\setlist{itemsep=2pt,topsep=5pt}
\newtheorem{theorem}{Theorem}[section]
\newtheorem{lemma}[theorem]{Lemma}

\theoremstyle{remark}
\newtheorem{remark}[theorem]{Remark}
\newcommand{\h}{\mathsf h}
\newcommand\unit{\hbox{\rm 1\kern-2.8truept l}}
\newcommand{\D}{\mathcal D_{\mathrm{fin}}}

\newcommand{\Dom}{\operatorname{Dom}}
\newcommand{\Rea}{\operatorname{Re}}
\newcommand{\Span}{\operatorname{span}_{\mathbb C}}
\newcommand{\norm}[1]{\left\lVert#1\right\rVert}
\newcommand{\ip}[2]{\left\langle#1,#2\right\rangle}
\newcommand{\Qtwo}{\mathscr Q_2}
\allowdisplaybreaks[2]
\title{On the domain of the effective non-Hermitian Hamiltonian in quantum markovian dynamics}
\author{Franco Fagnola$^{\footnotesize(1)}$, Zheng Li$^{\footnotesize(2)}$}
\date{ }
\begin{document}
\maketitle
\vspace{-1em}
\begin{abstract}
The GKLS representation of the unbounded generator of a quantum Markov semigroup
crucially depends on an unbounded operator $G$, generating a strongly continuous
contraction semigroup on the underlying Hilbert space, which is formally the sum
of a negative self-adjoint operator $G_0$ and $-\mathrm{i}H$ with $H$ symmetric.

In many situations it is useful to know the relationships between the domains of the three operators,
in particular whether $\Dom(G)=\Dom(G_0)\cap\Dom(H)$ when both $G_0$ and $H$ are unbounded.
In this note we show that this happens when $G_0$ and $H$ are self-adjoint quadratic polynomials
in annihilation and creation operators in a finite-mode bosonic Fock space and $G_0\leq0$ 
or assuming a certain half bound on the commutator of $G_0$ and $H$.

A counterexample shows that the conclusion does not hold in general situations.
\end{abstract}

\section{Introduction}
The celebrated Gorini-Kossakowski-Sudharshan-Lindblad (GKLS) representation of generators of
norm continuous quantum Markov semigroups (QMS)
on $\mathcal{B}(\mathsf{h})$,
\[
\mathcal{L}(x)= \mathrm{i}[H,x] -\frac{1}{2}\sum_{\ell}\left(L_\ell^*L_\ell x - 2L^*_\ell x\, L_\ell + x\,L_\ell^*L_\ell\right)
\]
with $H,L_\ell$ bounded operators on $\mathsf{h}$, $H=H^*$, is highly useful because it allows for the separation of
the Hamiltonian and dissipative components in the evolution of a quantum system; furthermore, it enables algebraic
calculations to be performed on operators acting on a Hilbert space, rather than on superoperators in a Banach space.

In its natural generalization to unbounded operators $H$ and $L_\ell$ (see \cite{BoFaQu,CheFag,Dav77,FF-Proy} and the references
therein) one starts from the hypothesis that
these operators share a dense common domain and the closure of $G=-\mathrm{i}H-\sum_\ell L_\ell^*L_\ell/2$,
($\mathrm{i} G$ is also called the effective non-hermitian Hamiltonian in physics literature), generates a strongly
continuous semigroup of contractions on $\mathsf{h}$.
However, in the analysis of various models, it becomes useful to be able to separate again the self-adjoint and
anti-self-adjoint parts of $G$ making sure that the domains behave well, i.e., as one would desire,
$\Dom(G)=\Dom(G_0)\cap\Dom(H)$
where $G_0$ denotes a suitable self-adjoint non-positive extension of $-\sum_\ell L_\ell^*L_\ell/2$.

In this note we show that this happens when $G_0$ and $H$ are self-adjoint quadratic polynomials
in annihilation and creation operators in a finite-mode bosonic Fock space and $G_0\leq0$ without assuming any special
properties for $G_0$ and $H$, such as ellipticity, hypoellipticity or other special property of the semigroup generated
by $G$. Addressing the problem proves to be non-trivial.

A counterexample shows that the conclusion does not hold in general situations.

The paper is organised as follows: in Section \ref{sect:gaussian-statement} we prove the result (Theorem \ref{th:graph-norm})
for quadratic operators in a finite-mode Fock space, in Section \ref{sect:comm-cond} we establish a sufficient condition
based on a lower bound of the commutator of the selfadjoint and anti-self adjoint part. Finally, in Section \ref{sect:gen-case}
we present a counterexample showing that the identity $\Dom(G)=\Dom(G_0)\cap\Dom(H)$ does not hold in a general situation.

\section{Setting and statement}\label{sect:gaussian-statement}

Fix a positive integer $d$. Let
\[
 \h=\Gamma_s(\mathbb C^d)
     =\bigoplus_{n=0}^{\infty}(\mathbb C^d)^{\otimes_s n}
\]
be the symmetric (bosonic) Fock space; its zeroth summand is $\mathbb C$.
We use scalar products linear in the second variable and write $\norm{u}$ for
the Hilbert-space norm. The same notation denotes the operator norm of a
bounded operator. The symbol $\unit$ denotes the identity operator.
For $\alpha=(\alpha_1,\ldots,\alpha_d)\in\mathbb{N}^d$, let $|\alpha\rangle$ be the normalized occupation vector, and put
\[
 \D=\Span\{\,|\alpha\rangle\,\mid\,\alpha\in\mathbb{N}^d\,\}.
\]
Thus each vector in $\D$ is a finite linear combination. If $e_j$ is the
$j$th coordinate vector, the annihilation and creation operators satisfy
\[
 a_j|\alpha\rangle=\sqrt{\alpha_j}\,|\alpha-e_j\rangle,
 \qquad
 a_j^\dagger |\alpha\rangle=\sqrt{\alpha_j+1}\,|\alpha+e_j\rangle
 \quad(1\leq j\leq d),
\]
where the first expression is zero if $\alpha_j=0$. On $\D$ their canonical
commutation relations are
\[
 [a_j,a_k^\dagger]=\delta_{jk}\unit,\qquad [a_j,a_k]=[a_j^\dagger ,a_k^\dagger ]=0.
\]
Here $[X,Y]=XY-YX$, and $\delta_{jk}$ is $1$ if $j=k$ and $0$ otherwise.

Let $\Qtwo$ be the finite-dimensional complex vector space of all normally
ordered polynomials of total degree at most two, acting on $\D$:
\begin{equation}
 B=\sum_{|\alpha|+|\beta|\leq2}b_{\alpha\beta}(a^\dagger)^\alpha a^\beta.
 \label{poly}
\end{equation}
Here $\alpha,\beta\in\mathbb{N}^d$,
$|\alpha|=\sum_{j=1}^d\alpha_j$,
$(a^\dagger)^\alpha=\prod_{j=1}^d(a_j^\dagger )^{\alpha_j}$,
$a^\beta=\prod_{j=1}^d a_j^{\beta_j}$, and
$b_{\alpha\beta}\in\mathbb C$ are fixed coefficients. In particular,
``quadratic'' below always allows linear and constant terms.

Suppose that $A$ and $H$ are the self-adjoint closures of symmetric members
of $\Qtwo$, and that $A\geq0$. A bar over an operator denotes its graph
closure; $\Dom(X)$ denotes the domain of an operator $X$. We use the same letters $A,H$
for their restrictions when doing polynomial algebra on $\D$. Define
\begin{equation}
 G_0=-A,\qquad G=-\overline{(A+\mathrm{i}H)|_{\D}}\, .
 \label{defG}
\end{equation}
It was shown in \cite[Appendix A, Theorem 4.1 and Proposition 4.2]{AFP}, applying the main result
of \cite[Theorem 2]{Jorgensen} that $G$  generates a strongly continuous contraction semigroup
$(P_t)_{t\geq 0}=(\mathrm{e}^{tG})_{t\geq 0}$ on $\mathsf{h}$. This generation result does not depend
on the sign of $H$. Therefore, it can be applied also to the closure $G^{-}$ of $-(A-\mathrm{i}H)|_{\D}$ that
generates another strongly continuous contraction semigroup $ (P^{-}_t)_{t\geq 0}$.

\begin{theorem}\label{th:graph-norm}
Under the above assumptions there is a finite constant $\kappa $ such that
\begin{equation}\label{eq:graph}
 \norm{Au}+\norm{Hu}\leq\kappa \bigl(\norm{Gu}+\norm{u}\bigr),
 \qquad u\in\D.
\end{equation}
Consequently
\begin{equation}\label{domain}
 \Dom(G)=\Dom(A)\cap\Dom(H)=\Dom(G_0)\cap\Dom(H),
 \qquad Gu=-Au-\mathrm{i}Hu.
\end{equation}
\end{theorem}

The sign condition is already forced by contraction generation if it was
not separately assumed: for $u\in\D$, dissipativity gives
$\ip{u}{G_0u}=\Rea\ip{u}{Gu}\leq0$, and the core property for $G_0$
extends this inequality to $\Dom(G_0)$.

Theorem \ref{th:graph-norm} applies to the common  choice for Gaussian QMSs
\begin{equation}
 A=\overline{\left.\frac12\sum_{\ell=1}^m L_\ell^*L_\ell\right|_{\D}}\, ,
 \qquad
 L_\ell=    \lambda_\ell\unit+   \sum_{j=1}^d(u_{\ell j}a_j+v_{\ell j}a_j^\dagger )\, .
 \label{noise}
\end{equation}
Here $m$ is a finite nonnegative integer and all coefficients
$\lambda_\ell,u_{\ell j},v_{\ell j}$ are complex; the adjoints in the
polynomial on $\D$ are the formal adjoints. The factor $1/2$ is a convention.
The proof needs only $A\geq0$, not the representation \eqref{noise}.

\begin{remark}
It is worth noting, at this point, that although the GKLS representation of a Gaussian generator is not unique
(see \cite{KRP}, Theorem 30.16) given that transforming the operators $L_\ell$ into $L'_\ell = \sum_k R_{\ell k} L_k$
via a unitary matrix $R$ does not, as is obvious, alter the operator $A$. Furthermore, by translating the operators
$L_\ell$ into $L'_\ell = L_\ell -\lambda_\ell\unit$ for a $m$-tuple of complex numbers $(\lambda_k)_{k=1}^m$,
the operator $G$ transforms into $G'=G+\sum_\ell\overline{\lambda}_\ell L_\ell -(\sum_{\ell}|\lambda_\ell|^2)\unit/2$,
which is found to have the same domain as $G$ due to the fundamental identity $\sum_{\ell}\norm{L_\ell u}^2=-2\operatorname{Re}\langle u, Gu\rangle$
satisfied by the GKLS forms of the Markovian generators. Indeed, this identity implies that operators $L_\ell$ are
relatively bounded with respect to $G$. The domain of the new Hamiltonian $H'$, however, may change.
\end{remark}

We begin the proof of Theorem \ref{th:graph-norm} by the following simple Lemma

\begin{lemma}\label{lem:loc-quadratic}
For all $t\geq 0$ we have $P^{-}_t = P^*_t$ and
\begin{equation}\label{eq:energy}
 \frac{\mathrm{d}}{\mathrm{d}t}\norm{P_t u}^2=-2\norm{A^{1/2}P_t u}^2
 \qquad\quad
 \frac{\mathrm{d}}{\mathrm{d}t}\norm{P_t^* v}^2=-2\norm{A^{1/2}P_t^*v}^2
\end{equation}
for all $u\in \operatorname{Dom}(G)$, $v\in \operatorname{Dom}(G^*)$\, .
\end{lemma}
\begin{proof}
For fixed $v\in\D$ and every $u\in\D$, $\ip{v}{Gu}=\ip{G^-v}{u}$. 
Since $\D$ is a core for $G$, 
this identity extends to $u\in\Dom(G)$, so $v\in\Dom(G^*)$ 
and $G^*v=G^-v$. 
Taking closures gives $G^-\subseteq G^*$. 
Both operators are maximal dissipative: 
$G^-$ by generation, 
and $G^*$ as the generator of 
the adjoint contraction semigroup. 
Hence $G^-=G^*$ and $P_t^-=P_t^*$.

To prove \eqref{eq:energy} note that, for all $u\in\D$ we have $-2\operatorname{Re}\langle u, G u \rangle = 2\Vert A^{1/2} u\Vert^2$.
Since $\D$ is a core for $G$, for graph approximants $u_n\in\D$ of $u\in\operatorname{Dom}(G)$, we have
\[
\norm{A^{1/2}(u_n-u_m)}^2
\leq\norm{u_n-u_m}\,\norm{G(u_n-u_m)}
\]
Therefore $u\in\operatorname{Dom}(A^{1/2})$ and the identity $-2\operatorname{Re}\langle u, G u \rangle = 2\Vert A^{1/2} u\Vert^2$
holds for $u\in \operatorname{Dom}(G)$.  It follows that
\[
 \frac{\mathrm{d}}{\mathrm{d}t}\norm{P_t u}^2
 =2\operatorname{Re}\langle P_t u, G P_t u \rangle = -2\norm{A^{1/2}P_t u}^2.
\]
The identity for $P^*_t$ can be established in the same way.
\end{proof}

The following lemma is inspired by the polynomial reproducing kernel
and inverse-moment-matrix construction in \cite{Simon} Sect. 2.

\begin{lemma}\label{lem:average}
There exists $\tau>0$ such that, for all $f\in\D$ and $0<t<\tau$, we have
\begin{equation}\label{eq:average}
 AP_t f =\int_{1/4}^{3/4} k_t(s)P_{ts}A P_{(1-s)t}f\,\mathrm{d}s
\end{equation}
where $(t,s)\mapsto k_t(s)$ is smooth and bounded on $[0,\tau]\times [1/4,3/4]$.
\end{lemma}

\begin{proof}

The canonical commutation relations give $[\Qtwo,\Qtwo]\subset\Qtwo$.
Define the derivation on $\Qtwo$ $\delta_G(B)=[B,G]$ algebraically on $\D$, and
$\mathcal{V}=\operatorname{span}\{A,\delta_G(A),\delta_G^2 (A),\ldots\}$.
If $A=0$ the result is immediate; otherwise, with $r=\dim \mathcal{V}$,
$A,\delta_G(A),\ldots,\delta_G^{r-1}(A)$ is a basis of $\mathcal{V}$.
Clearly, in this basis, by finite dimensionality of $\mathcal{V}$, there exists an $r\times r$ matrix $\Delta$ such that
\[
\left[\begin{array}{c} \delta^r_G(A) \\ \dots \\ \delta^{2r-1}_G(A)\end{array}\right]
= \Delta\left[\begin{array}{c} A \\ \dots \\ \delta^{r-1}_G(A)\end{array}\right]
\]
and so, for all $k>0$,
\[
\left[\begin{array}{c} \delta^{kr}_G(A) \\ \dots \\ \delta^{(k+1)r-1}_G(A)\end{array}\right]
= \Delta^k\left[\begin{array}{c} A \\ \dots \\ \delta^{r-1}_G(A)\end{array}\right].
\]
It follows that
\begin{align*}
 \mathrm{e}^{-ts\,\delta_G}A  & = \sum_{m\geq 0} \frac{(-ts)^m}{m!} \delta_G^m(A)\\
    & = \sum_{i=0}^{r-1} \sum_{k=0}^\infty\frac{(-ts)^{kr+i}}{(kr+i)!}\sum_{j=0}^{r-1} (\Delta^k)_{ij}  \delta_G^j(A)\\
    & =  \sum_{j=0}^{r-1}\frac{(-ts)^{j}}{j!} \delta_G^j(A)
    +\sum_{i=0}^{r-1} \sum_{k=1}^\infty\frac{(-ts)^{kr+i}}{(kr+i)!}\sum_{j=0}^{r-1} (\Delta^k)_{ij}  \delta_G^j(A)\\
    & =  \sum_{j=0}^{r-1}\frac{(-ts)^{j}}{j!} \delta_G^j(A)
    +\sum_{j=0}^{r-1} t^j\left(\sum_{i=0}^{r-1} \sum_{k=1}^\infty\frac{t^{kr+i-j}(-s)^{kr+i}}{(kr+i)!}(\Delta^k)_{ij}\right)  \delta_G^j(A)\, .
\end{align*}
Therefore, for  $s\in \mathcal I=[1/4,3/4]$
\[
 \mathrm{e}^{-ts\,\delta_G}A=\sum_{j=0}^{r-1}t^j h_j(t,s)\delta_G^j(A)
\]
where the coordinate functions $ h_j$ are continuous at $t=0$, uniformly for
$s\in\mathcal I$, and $h_j(0,s)=(-s)^j/j!$.

Let $M(t)$ be the $r\times r$ matrix with entries
\[
 M(t)_{jk}=\int_{\mathcal I} h_j(t,s)s^k\,\mathrm{d}s,
 \qquad 0\leq j,k<r.
\]
Its limit at zero is
\[
M(0)_{jk}= \frac{(-1)^j}{j!} \int_{\mathcal I}  s^j s^k \,\mathrm{d}s
 \qquad 0\leq j,k<r
\]
and is invertible as the product of the invertible diagonal matrix $\operatorname{diag}((-1)^j/j!) $ and
another matrix which is strictly positive definite by
\[
\sum_{j,k}\overline{w}_j  \int_{\mathcal I}  s^j s^k \,\mathrm{d}s\, w_k
= \int_{\mathcal I}  \Big|\sum_k w_k s^k\Big|^2 ds \qquad \forall w\in\mathbb{C}^r\, .
\]
Choose $T_0>0$ so that $M(t)$ is invertible for $0\leq t\leq T_0$, and let
$c(t)=M(t)^{-1}[1,0,\ldots,0]^{\mathsf T}$ and $k_t(s)=\sum_{k=0}^{r-1}c_k(t)s^k$.
Note that $(t,s)\mapsto k_t(s)$ is a complex valued smooth function on $[0,\tau]\times \mathcal{I}$
and $\int_{\mathcal I} k_t(s)h_j(t,s)\,ds=\delta_{j0}$.
Moreover, there is $K<\infty$ such that
\begin{equation}\label{eq:reproduce}
 \sup_{0\leq t\leq T_0,\ s\in\mathcal I}|k_t(s)|\leq K,\qquad
 A=\int_{\mathcal I} k_t(s)e^{-ts\,\delta_G}A\,ds.
\end{equation}

For $f\in\D$ and $0<t<\tau$, consider $F_s(B)=P_s B P_{t-s}f$, $B\in \mathcal{V}$, $0\leq s\leq t$.
Since the domain $\operatorname{Dom}(N^{n/2})$, where $N=\sum_{j=1}^d a^\dagger_j a_j$ is the number operator,  
is $P_t$-invariant for all $n \ge 0$ and $t \ge 0$ and there exist
constants $c_n > 0$ such that
\begin{equation*}
\|(N + \unit)^{n/2} P_t \psi\|^2 \le e^{4 c_n t} \|(N + \unit)^{n/2} \psi\|^2
\end{equation*}
for all $\psi \in \operatorname{Dom}(N^{n/2})$ (see \cite{FF-DP-IDAQP-2022} Lemma 6),
$s\mapsto B P_{t-s}f$ is continuously differentiable in the Hilbert space, belongs to $\Dom(G)$, and has continuous $G$-image.
The semigroup differentiation rule gives $F_s'(B)=-F_s(\delta_G(B))$. This is a finite system of linear ordinary
differential equations with initial values $F_0(B)=B P_tf$. Thus
\[
 (\mathrm{e}^{-ts\,\delta_G}B)P_tf=P_{ts}B P_{(1-s)t}f,
 \qquad s\in\mathcal{I},\quad B\in \mathcal{V}.
\]
Combining this with \eqref{eq:reproduce} yields \eqref{eq:average}.
\end{proof}

We are now in a position to prove Theorem \ref{th:graph-norm}.

\begin{proof} {\it (of Theorem \ref{th:graph-norm})}
Define, for arbitrary Hilbert-space vectors,
\[
 d_u(t)=\norm{P_{t/4}u}^2-\norm{P_{3t/4}u}^2,
 \qquad
 d_v^*(t)=\norm{P_{t/4}^*v}^2-\norm{P_{3t/4}^*v}^2.
\]
These quantities are nonnegative by contractivity. Pair
\eqref{eq:average} with $v\in\D$. Self-adjoint positivity of $A$, the
Cauchy--Schwarz inequality on $\mathcal I$, and \eqref{eq:energy} give
\begin{align*}
 |\ip{v}{AP_tu}|
 &\leq K\left(\int_{\mathcal I}\norm{A^{1/2}P_{ts}^*v}^2ds\right)^{1/2}
          \left(\int_{\mathcal I}\norm{A^{1/2}P_{(1-s)t}u}^2ds\right)^{1/2}\\
 &=\frac{K}{2t}\sqrt{d_u(t)d_v^*(t)}.
\end{align*}
By density the estimate holds for every $v\in\mathsf h$, still with
$u\in\D$. Choose $T>0$ such that all previous formulas hold for
$0<t\leq T$. In particular
\begin{equation}\label{fixed}
 \norm{AP_Tu}\leq\frac{K}{2T}\norm{u},\qquad u\in\D.
\end{equation}
Let us now use a truncated version of the well-known Frullani identity \cite{Frullani}.
For $u\in\h$, a change of variable, first with a lower cutoff $\varepsilon>0$, gives
\begin{align*}
 \int_\varepsilon^T d_u(t)\,\frac{\mathrm{d}t}{t}
 &=\int_{\varepsilon/4}^{T/4}\norm{P_su}^2\,\frac{\mathrm{d}s}{s}
   -\int_{3\varepsilon/4}^{3T/4}\norm{P_s u}^2\,\frac{\mathrm{d}s}{s}\\
 &=\int_{\varepsilon/4}^{3\varepsilon/4}\norm{P_s u}^2\,\frac{\mathrm{d}s}{s}
   -\int_{T/4}^{3T/4}\norm{P_s u}^2\,\frac{\mathrm{d}s}{s}
\end{align*}
for $0<\varepsilon<T/3$. Strong continuity at zero and then
$\varepsilon\downarrow0$ show
\begin{equation}\label{eq:telescope}
 \int_0^T d_u(t)\,\frac{\mathrm{d}t}{t}
 = (\log3)\norm{u}^2
   -\int_{T/4}^{3T/4}\norm{P_s u}^2\,\frac{\mathrm{d}s}{s}
 \leq(\log3)\norm{u}^2.
\end{equation}
The same inequality holds for the adjoint semigroup, which is strongly continuous
on a Hilbert space. Therefore Cauchy--Schwarz for $dt/t$ yields
\begin{equation}\label{eq:integrated}
 \int_0^T|\ip{v}{AP_t u}|\,\mathrm{d}t
 \leq\frac{K\log3}{2}\norm{u}\norm{v},
 \qquad u\in\D,\quad v\in\mathsf h.
\end{equation}
For $u\in\D$ we also have $G u\in\D$ and, by \cite{FF-DP-IDAQP-2022} Lemma 6 we can differentiate
$t\mapsto AP_tu$ finding
\[
 Au=AP_Tu-\int_0^T AP_tGu\,\mathrm{d}t.
\]
Taking the scalar product with $v$, applying \eqref{fixed} and \eqref{eq:integrated}, and taking
the supremum over unit vectors $v$ we obtain
\[
 \norm{Au}\leq\frac{K}{2T}\norm{u}
            +\frac{K\log3}{2}\norm{Gu},\qquad u\in\D.
\]
Since $\mathrm{i}Hu=-Gu-Au$ we have
\begin{align*}
 \norm{Au}+\norm{Hu} & \leq  \norm{Gu}+2\norm{Au} \\
 & \leq (1+K\log 3)\norm{Gu} + KT^{-1}\norm{u}
\end{align*}
and the claimed graph estimate \eqref{eq:graph} readily follows. This implies that
$\operatorname{Dom}(G)\subseteq\operatorname{Dom}(A)\cap\operatorname{Dom}(H)$. Indeed, take
$(u_n)_{n\geq 1}\subseteq\D$ converging to $u\in\Dom(G)$ in the graph norm of $G$ and apply
\eqref{eq:graph} to $u_n-u_m$.  The sequences $(Au_n)_{n\geq 1}$ and $(Hu_n)_{n\geq 1}$ are Cauchy;
closedness of $A$ and $H$ gives
\[
u\in\Dom(A)\cap\Dom(H),\qquad Gu=-Au-\mathrm{i}Hu.
\]
To prove the reverse inclusion $\operatorname{Dom}(A)\cap\operatorname{Dom}(H)\subseteq\operatorname{Dom}(G)$
consider the operator $X=-A-\mathrm iH$ on $\operatorname{Dom}(A)\cap\operatorname{Dom}(H)$. We have $G\subseteq X$,
and $X$ is dissipative because $A\geq0$. Since $G$ is maximal dissipative because it is dissipative and generates
a strongly continuous contraction semigroup, then $X=G$. Explicitly, for $u\in\operatorname{Dom}(X)$
we can choose $v\in\operatorname{Dom}(G)$ with $(\unit-G)v=(\unit-X)u$. Then $w=u-v$ satisfies
\[
0=\operatorname{Re}\ip{w}{(\unit-X)w}
=\norm{w}^2+\ip{w}{Aw}\geq\norm{w}^2,
\]
so $u=v$. This completes the proof.
\end{proof}

\begin{remark}
We believe that an alternative proof can be obtained using subelliptic estimates for quadratic differential operators
(see \cite{AB,HPV}). Here, we have opted for a proof that is as direct as possible, employing techniques for quadratic
operators in Fock spaces—an approach with which we are more familiar.
\end{remark}

\section{A commutator condition}\label{sect:comm-cond}

We present now a general result assuming a certain lower bound on the commutator of
$A$ and $H$ on a general Hilbert space $\mathsf{k}$. Throughout this section we assume that:
\begin{itemize}
  \item[(i)] The operators $A$ and $H$ are self-adjoint and $A\geq 0$.
  \item[(ii)] There exists a dense domain $D_o$ which is an essential domain for both $A$ and $H$ such that
  $A\, D_o\subseteq \operatorname{Dom}(H)$, $H\, D_o \subseteq \operatorname{Dom}(A)$.
  \item[(iii)] The closure $G$ of $-(A+\mathrm{i}H)|_{D_o}$ generates a strongly continuous semigroup
  on $\mathsf{k}$.
  \item[(iv)] There exist constants  $0\leq c_1<1, c_2>0$ such that, for all $u\in D_o$,
  \[
 \left\langle u,\mathrm{i}[A,H]u\right\rangle
\geq -c_1 \left(\left\Vert Au \right\Vert^2 + \left\Vert Hu \right\Vert^2\right) - c_2 \left\Vert u \right\Vert^2\, .
\]
\end{itemize}

\begin{theorem}\label{th:commutator-lb}
 Under the above conditions (i) -- (iv) we have $\operatorname{Dom}(G)=\operatorname{Dom}(A)\cap\operatorname{Dom}(H)$.
\end{theorem}

\begin{proof} Because of (iv), we have
\begin{align*}
\norm{Gu}^2 & = \norm{Au}^2 + \mathrm{i}\ip{u}{[A,H]u} + \norm{Hu}^2 \\
& \geq  (1-c_1) \left( \norm{Au}^2  + \norm{Hu}^2\right) - c_2 \left\Vert u \right\Vert^2.
\end{align*}
for all $u\in D_o$. Therefore
\begin{align*}
\sqrt{1-c_1}\left( \norm{Au}   + \norm{Hu} \right)
&  \leq \left( 2(1-c_1)\left(\norm{Au}^2  + \norm{Hu}^2\right)\right)^{1/2} \\
& \leq \left(2\norm{Gu}^2+2c_2\norm{u}^2\right)^{1/2}\\
& \leq \sqrt{2}\left(\norm{Gu}+\sqrt{c_2}\norm{u}\right)
\end{align*}
supplies the necessary inequality for showing that $\operatorname{Dom}(G)\subseteq\operatorname{Dom}(A)\cap \operatorname{Dom}(H)$
and $G\subseteq -A-\mathrm{i} H$ on $\operatorname{Dom}(A)\cap \operatorname{Dom}(H)$.  
The conclusion follows as the last step of the proof of Theorem \ref{th:graph-norm}.
\end{proof}

\section{The question for general self-adjoint operators}\label{sect:gen-case}

In this section we show that the corresponding abstract assertion is false, even with a common
invariant core and strongly continuous contraction semigroup generation for both $-(A\pm\mathrm{i}H)$.
We present a counterexample for abstract direct-sum operators.

Let us consider the Hilbert space $\mathsf{k}=\bigoplus_{n=1}^{\infty}\mathbb C^2$, with squared norm
$\norm{x}^2=\sum_{n\geq 1}\norm{x_n}_{\mathbb C^2}^2$ for $x=(x_n)_{n\geq1}$.
Define the Hermitian matrices
\[
 A_n=\begin{bmatrix}n^2&n\\n&1\end{bmatrix}=\begin{bmatrix}n\\1\end{bmatrix}[n\, 1]\, ,\qquad
 H_n=\begin{bmatrix}0&\mathrm{i}n\\-\mathrm{i}n&0\end{bmatrix}\, ,
 \qquad n\in\mathbb N^*=\{\,1,2,\dots\,\}.
\]
Let $A_{\mathrm{ex}}$ and $H_{\mathrm{ex}}$ be their self-adjoint direct sums with natural domains
\[
 \Dom(A_{\mathrm{ex}})=
 \left\{x\in\mathsf{k}\,\Big|\,\sum_{n\geq 1}\norm{A_nx_n}^2<\infty\right\},
\]
and the analogous definition with $H_n$ for $\Dom(H_{\mathrm{ex}})$.
Vectors with finitely many nonzero blocks form a common invariant core
$\mathcal D_{\mathrm{ex}}$, as follows by truncating the square sums.
Put $G_{0,\mathrm{ex}}=-A_{\mathrm{ex}}$ and
\[
 C_n=A_n+\mathrm{i}H_n=\begin{bmatrix}n^2&0\\2n&1\end{bmatrix},\qquad
 C_n^*=A_n-\mathrm{i}H_n=\begin{bmatrix}n^2&2n\\0&1\end{bmatrix}.
\]
Both matrices have nonnegative Hermitian part $A_n$, so
$\norm{\mathrm{e}^{-tC_n}}\leq1$ and $\norm{\mathrm{e}^{-tC_n^*}}\leq1$ for $t\geq0$,
by differentiation of squared norms. The direct sums of these
matrix exponentials are strongly continuous contraction semigroups
on $\mathsf{k}$. Strong continuity follows on finite-block vectors
first, and then by density and contractivity.

Let $C_{\mathrm{ex}}=\bigoplus_n C_n$ with
\[
 \Dom(C_{\mathrm{ex}})=
 \left\{x\in\mathsf{k}\,\Big|\,\sum_{n\geq 1}\norm{C_nx_n}^2<\infty\right\}.
\]
The two generators are $-C_{\mathrm{ex}}$ and $-C_{\mathrm{ex}}^*$.
To verify the first domain description, coordinatewise differentiation
shows that any vector in the generator domain must have
$(C_nx_n)_n\in\mathsf{k}$. Conversely, for such a vector, we can write
\[
 \left(\mathrm{e}^{-tC_n}-\unit\right)x_n=-\int_0^t \mathrm{e}^{-sC_n}C_nx_n\,\mathrm{d}s
\]
and the summable bound $\norm{C_nx_n}^2$ allows us to pass to the derivative
with respect to the norm of $\mathsf{k}$. The adjoint case is identical. Finite-block truncations
are graph approximants for both direct sums. Thus these generators
are exactly the closures of
$(G_{0,\mathrm{ex}}-\mathrm{i}H_{\mathrm{ex}})|_{\mathcal D_{\mathrm{ex}}}$ and
$(G_{0,\mathrm{ex}}+\mathrm{i}H_{\mathrm{ex}})|_{\mathcal D_{\mathrm{ex}}}$.

Now choose $x_n=[0,1/n]^{\mathsf T}$, where $\mathsf T$ denotes transpose. Then $x=(x_n)_n\in\mathsf{k}$ and
\[
 C_nx_n=\begin{bmatrix} 0 \\ 1/n \end{bmatrix},\qquad
 A_nx_n=\begin{bmatrix} 1 \\ 1/n \end{bmatrix},\qquad
 H_nx_n=\begin{bmatrix} \mathrm{i} \\ 0 \end{bmatrix}.
\]
Consequently $x\in\Dom(C_{\mathrm{ex}})$ but
$x\notin\Dom(A_{\mathrm{ex}})$ and
$x\notin\Dom(H_{\mathrm{ex}})$. The desired intersection identity
therefore fails. In fact $C_n^*x_n=[2,1/n]^{\mathsf T}$, so even the
two generator domains differ.

Note that this example is consistent with Section 3 because on the second unit vector of block $n$,
one has
\[
\ip{e_2}{\mathrm i[A_n,H_n]e_2}=-2n^2,
\qquad \norm{A_ne_2}^2+\norm{H_ne_2}^2=2n^2+1,
\]
so the uniform lower bound with $c_1<1$ cannot hold.

\begin{remark}
Throughout our analysis, contraction generation refers to the
\emph{closures} of $-(A\pm\mathrm{i}H)$ on a core. If instead an algebraic
sum on $\Dom(A)\cap\Dom(H)$ is assumed to be a generator already,
its domain is that intersection by definition. But then the problem
is to check this property which is the first key step for the
construction of QMS by the minimal semigroup method (see \cite{Dav77,CheFag,FF-Proy}).
\end{remark}

\begin{remark}{\it (singular quantum models).}
The quantum Az\'ema construction of Chebotarev--Fagnola \cite{CF95}, Section 2, illustrates a related domain difficulty
outside the self-adjoint Hamiltonian setting. In this model the effective generator is the original paper's operator $G^*$,
here denoted by $K$ to avoid confusion. For $\beta\in\mathbb R -\{0,-1\}$, one considers operators on
$D_0=C_c^\infty(\mathbb R\setminus\{0\})$
\[
K u=\frac{u'}{\beta x}-\frac{\beta+1}{2\beta^2x^2}u,
\qquad Mu=\frac{u}{\beta x},\qquad L=SM,
\]
where $S$ is unitary so that, in the notation of this note,
\[
A=\tfrac12M^2,\qquad Au=\frac{u}{2\beta^2x^2},\qquad
K=-A-\mathrm iH_0,
\quad H_0=\frac{\mathrm i}{\beta}\left(x^{-1}\partial_x-\frac1{2x^2}\right).
\]
The unitary change of variables
\[
(Uu)_\pm(y)=(2y)^{-1/4}u(\pm\sqrt{2y}),\qquad y>0,
\]
transforms $H_0$ into two copies of $(\mathrm{i}/\beta)\partial_y$ with the same orientation.
The closure has domain $H_0^1(0,\infty)^{\oplus2}$ and the adjoint has domain $H^1(0,\infty)^{\oplus2}$.
With $n_\pm=\dim\ker(H_0^*\mp\mathrm i)$, the deficiency solutions are $e^{\beta y}$ and $e^{-\beta y}$ respectively, so
\[
(n_+,n_-)=
\begin{cases}(0,2),&\beta>0,\\(2,0),&\beta<0.\end{cases}
\]
Thus there is no self-adjoint extension on the stated Hilbert space, even if one allows coupling at the two endpoints.
\end{remark}

\section*{Use of AI tools}
During the preparation of this manuscript, the authors used ChatGPT to assist with language editing,
obtaining exploratory suggestions regarding mathematical proofs and checking intermediate versions.
All statements, proofs, references, and conclusions were independently verified, reviewed,
and approved by the authors, who assume full responsibility for the content of the manuscript.

\noindent{\scriptsize $^{(1)}$ Franco Fagnola, Dipartimento di Matematica, Politecnico di Milano, Piazza Leonardo da Vinci 32, Milano, 20133, Italy. Email: franco.fagnola@polimi.it}

\noindent{\scriptsize $^{(2)}$ Zheng Li, School of Mathematics and Statistics, Central South University, Changsha, 410083, China.
Email: zheng.li@csu.edu.cn}

\end{document}